\documentclass[preprint,showpacs,preprintnumbers,amsmath,amssymb]{revtex4}

\usepackage{epsfig}
\usepackage{graphicx}
\usepackage{dcolumn}
\usepackage{bm}

\def\beq{\begin{equation}}
\def\eeq#1{\label{#1}\end{equation}}
\def\eeqn{\end{equation}}
\newcommand\iden{\leavevmode\hbox{\small1\normalsize\kern-.33em1}}

\let\jnfont=\rm
\def\NPB#1,{{\jnfont Nucl.\ Phys.\ B }{\bf #1},}
\def\PLB#1,{{\jnfont Phys.\ Lett.\ B }{\bf #1},}
\def\EPJC#1,{{\jnfont Eur.\ Phys.\ Jour.\ C }{\bf #1},}
\def\PRD#1,{{\jnfont Phys.\ Rev.\ D }{\bf #1},}
\def\PRL#1,{{\jnfont Phys.\ Rev.\ Lett.\ }{\bf #1},}
\def\MPLA#1,{{\jnfont Mod.\ Phys.\ Lett.\ A }{\bf #1},}
\def\JPG#1,{{\jnfont J.\ Phys.\ G }{\bf #1},}
\def\CTP#1,{{\jnfont Commun.\ Theor.\ Phys.\ }{\bf #1},}
\def\JHEP#1,{{\jnfont JHEP \ }{\bf #1},}
\def\NPPS#1,{{\jnfont Nucl.\ Phys.\ Proc.\ Suppl.\ }{\bf #1},}
\def\CPC#1,{{\jnfont Computl.\ Phys.\ Commun.\ }{\bf #1},}

\begin{document}

\preprint{\parbox{1.2in}{\noindent  arXiv:0801.0210}}

\title{\ \\[10mm]  Probing New Physics from Top Quark FCNC Processes at LHC: \\ A Mini Review}

\author{\ \\[2mm] Jin Min Yang \\ ~}

\affiliation{Institute of Theoretical Physics, Academia Sinica,
             Beijing 100080, China \vspace*{1.5cm}}

\begin{abstract}
Since the top quark FCNC processes are extremely suppressed
in the Standard Model (SM) but could be greatly enhanced
in some new physics models, they could serve as a smoking gun for
new physics hunting at the LHC.
In this brief review we summarize the new physics predictions
for various top quark FCNC processes at the LHC by focusing on
two typical models: the minimal supersymmetric model (MSSM) and
the topcolor-assisted technicolor (TC2) model.
The conclusion is: (1) Both new physics models can greatly
enhance the SM predictions by several orders;
(2) The TC2 model allows for largest enhancement, and for each
channel the maximal prediction is much larger than in the MSSM;
(3) Compared with the $3\sigma$ sensitivity at the LHC, only a
couple of channels are accessible for the MSSM
while most channels are accessible for the TC2 model.
\vspace*{.3cm}

\vspace*{1cm}

\end{abstract}

\pacs{14.65.Ha, 14.80.Ly, 11.30.Hv\\ ~ \\ to appear in Int. J. Mod. Phys. A}

\maketitle

\section{Introduction}
With the forthcoming experiments at the Large Hadron Collider (LHC)
the elementary particle physics will come to a critical
crossroad: the discovery of new physics at TeV scale will bring a new
highlight and lift the curtain on a new exciting era while the undiscovery
of new physics will give a body blow to particle physics.
Among various speculations of new physics at TeV scale, there are
two typical directions: one is supersymmetry, which is a weak coupling theory
with fundamental scalars and can fancily push up the energy scale to ultimately
realize grand unification, and the other is the theory of dynamical symmetry
breaking (like technicolor) without fundamental scalars.
Undoubtedly, these beautiful theories will soon be put to the sword of the LHC.

For the probe of new physics at the high energy colliders like the LHC, there
are two ways: one is through detecting the direct production of new particles
and the other is through unraveling the quantum effects of new physics in some
sensitive and well-measured processes. These two aspects can be complementary
and offer a consistent check for a framework of new physics. If the collider
energy is high enough above the mass threshold of the new particles, studying 
the direct production of new particles plays the dominant role; while for the
collider energy not high enough to surpass the mass threshold of new particles,
disentangling the quantum effects of new particles will be the only way of peeking
at the hints of new physics.
Therefore, a collider with relatively low energy but high measuring precision
will serve as a telescope for looking at the new physics at much higher energy
scales.

Given the importance of quantum effects in probing new physics and the uncertainty 
of new physics scale, we should
seriously examine some LHC processes which are sensitive to new physics.
As the heaviest fermion in the SM, the top quark is speculated
to be a sensitive probe of new physics \cite{top-review}.
Due to the small statistics of the experiments at the Fermilab Tevatron collider,
so far the top quark properties have not been precisely measured and
there remained a plenty of room for new physics effects in top
quark processes. Since the LHC will be a top quark factory and allow to scrutinize
the top quark nature, unraveling new physics effects in various top quark
processes will be an intriguing channel for testing new physics models.

One typical property of the top quark in the SM is its
extremely small flavor-changing neutral-current (FCNC)
interactions  \cite{tcvh-sm} due to the Glashow-Iliopoulos-Maiani (GIM)
mechanism. This will make the observation of any
FCNC top quark process a smoking gun for new physics beyond the SM.
So far numerous studies \cite{top-fcnc-review} have been
performed and have shown that the FCNC top quark interactions
can be significantly enhanced in various new physics models.
Due to the fact that different models predicts different orders
of enhancement, the measurements of these FCNC top quark
processes at the LHC will not only shed some light on new
physics but also may possibly give some favor or unfavor
information for a specified model.
In this review we will summarize the predictions in
the minimal supersymmetric model (MSSM) \cite{tcv-pptc-mssm,tcv-mssm,pptc-mssm}
and the topcolor-assisted technicolor (TC2) model \cite{tcv-tc2,pptc-tc2,pptc-tc2-other}.
Since these two models represent two opposite directions for
new physics, they in principle do not co-exist and, as shown
in this review, they predict quite different enhancements
for these  FCNC top quark processes.

\section{FCNC top quark processes in MSSM and TC2}
{\em In the SM}: Due to the GIM mechanism, the top quark FCNC interactions
are absent at tree-level and are extremely
suppressed at loop-level since such FCNC interactions are induced
by the $W$-boson charge-current CKM transitions involving down-type quarks
in the loops which are much lighter than the top quark.  The top quark FCNC
induced by the $W$-boson loops are dependent on the mass splitting of the
down-type quarks appearing in the loops.
Neglecting the masses of these down-type quarks or assuming their degeneracy,
then the one-loop induced  top quark FCNC interactions will vanish since the
$KM$ matrix are unitary and the $W$-boson couplings to fermions are universial.

{\em In the MSSM}: Although the top quark FCNC interactions are also
induced at loop level, they can be greatly enhanced relative to the
SM predictions. In addition to the $W$-boson loops,
there are four kinds of loops contributing to the top quark FCNC interactions.
The first type is charged Higgs loops whose contributions can be much larger
than the $W$-boson loops since the Yukawa couplings are proportional to fermion
masses and non-universal for the down type quarks appearing in the loops.
The second type is chargino loops whose contributions
can be much larger than the $W$-boson loops since the mass splitting between
the squarks in the loops may be significant and the Higgsino-component couplings
are non-universal Yukawa couplings.
The third type is gluino loops due to the flavor mixings between stops and other
up-type squarks (mainly scharms). Since such stop-scharm flavor mixings may be
significant, this kind of loops involving the strong coupling may be quite sizable
or dominant over other kinds of loops. The forth type is neutralino loops, also 
due to the flavor mixings between stops and other up-type squarks, which are 
usually smaller in magnitude than gluino or chargino loops.  

{\em In the TC2 model}:
Technicolor is a typical idea to dynamically break the electroweak symmetry.
But the original simple technicolor theory encounters enormous
difficulty in generating fermion masses (especially the heavy top quark
mass) and phenomenologically face the difficulty of passing through
the precision electroweak test.
The topcolor-assited technicolor (TC2) model \cite{TC2-model}
combines technicolor with topcolor, with the former being responsible for
electroweak symmetry breaking and the latter for generating large
top quark mass. This model so far survives current experiments
and will be put to the sword at the LHC.

The top quark FCNC interactions may be greatly enhanced in TC2 model for the following reasons.
(1) Topcolor is non-universal, only causing top-quark to condensate and only giving top quark mass
(a large portion $1-\epsilon_t$). The neutral top-pion has large Yukawa couplings to only top quark.
(2) ETC (extended technicolor) gives masses to all quarks and for top quark only a small portion
$\epsilon_t$ of mass is from ETC. ETC-pions have small Yukawa couplings to all quarks, and
for top quark the coupling is much weaker than the top-pion's.
(3) Since the top quark mass and thus the mass matrix of up-type quarks is composed of both
ETC and topcolor contributions, the diagonalization of the mass matrix of up-type quarks
cannot ensure the simultaneous diagonalization of both the top-pion's Yukwawa couplings in
topcolor sector and the ETC-pions' Yukawa couplings in ETC sector.
Thus, after the diagonalization of the mass matrix of up-type quarks, the top-pion in topcolor
sector will have tree-level FCNC Yukwawa  couplings for the top quark.
This is in contrast to the SM Higgs boson whose couplings with the fermions have no
FCNC at tree-level because all fermion masses are from the couplings of only one Higgs doublet
and the diagonalization of the fermion mass matrix (given by the Yukawa coupling matrix times
a constant vev $v$) can simultaneously ensure the
diagonalization of the Yukawa coupling matrix.

In Table 1 we summarize the maximal predictions for five FCNC top quark decay modes
in the MSSM and TC2. The SM predictions are far below the LHC sensitivity and not
listed here. The MSSM maximal predictions \cite{tcv-pptc-mssm} were obtained from 
a scan over the parameter space by considering all current experimental constraints, 
such as the experimental bounds on squark and Higgs boson masses, the
precision measurements of $W$-boson mass and the effective weak
mixing angle as well as the experimental data on $B_s-\bar{B}_s$
mixing and $b \to s \gamma$.

\begin{center}
\null \noindent{\small Table 1: Maximal predictions for the branching ratios of 
                       FCNC top quark \\ ~~~~~~decays and
                       production cross sections (hadronic) at the LHC.}
\vspace{.3cm}

\doublerulesep 1.5pt \tabcolsep 0.2in
\begin{tabular}{llll} 
\hline \hline
             ~ & MSSM & TC2 & LHC $3\sigma$ sensitivity \\
$t\to cZ$
       & $1.8 \times 10^{-6}$ \cite{tcv-pptc-mssm} & $O(10^{-4})$ \cite{tcv-tc2}
       & $3.6 \times 10^{-5}$ \cite{lhc-3sigma} \\
$t\to c\gamma$
       & $5.2 \times 10^{-7}$ \cite{tcv-pptc-mssm} & $O(10^{-6})$ \cite{tcv-tc2}
       & $1.2 \times 10^{-5}$ \cite{lhc-3sigma}\\
$t\to ch$
       & $6.0 \times 10^{-5}$ \cite{tcv-pptc-mssm} & $O(10^{-1})$ \cite{tcv-tc2}
       & $5.8 \times 10^{-5}$ \cite{lhc-3sigma} \\
$t\to cg$
       &  $3.2 \times 10^{-5}$ \cite{tcv-pptc-mssm}  & $O(10^{-3})$ \cite{tcv-tc2}
       &  \\
$t\to cgg$
       & $3.5 \times 10^{-5}$ \cite{tcv-pptc-mssm} & $O(10^{-3})$ \cite{tcv-tc2}
       & \\ \hline
$gg \to t\bar{c}$
       & 700 fb \cite{tcv-pptc-mssm}& 30 pb \cite{pptc-tc2}
       & 1500 fb \cite{lhc-3sigma}\\
$cg \to t$
       &  950 fb \cite{tcv-pptc-mssm}& 1.5 pb \cite{pptc-tc2}
       & 800 fb \cite{lhc-3sigma} \\
$cg \to tg$
       & 520 fb \cite{tcv-pptc-mssm}& 3 pb \cite{pptc-tc2}
       & 1500 fb \cite{lhc-3sigma}\\
$cg \to t \gamma$
       &  1.8 fb  \cite{tcv-pptc-mssm}& 20 fb \cite{pptc-tc2}
       & 5 fb \cite{lhc-3sigma} \\
$cg \to tZ$
       & 5.7 fb  \cite{tcv-pptc-mssm}& 100 fb  \cite{pptc-tc2}
       & 35 fb \cite{lhc-3sigma}\\
$cg \to th$ &  24 fb \cite{tcv-pptc-mssm}& 1 pb \cite{pptc-tc2}
            & 200 fb \cite{lhc-3sigma}\\
\hline \hline
\end{tabular}
\end{center}

\section{Conclusion}
From Table 1 we draw the conclusion:
(1) Both new physics models can greatly
enhance the SM predictions by several orders;
(2) The TC2 model allows for largest enhancement, and for each
channel the maximal prediction is much larger than in the MSSM;
(3) Compared with the $3\sigma$ sensitivity of the LHC, only a
couple of channels are accessible in the MSSM
while most channels are accessible in the TC2 model.

\section*{Acknowledgments}
This work is supported in part by
the National Natural Science
Foundation of China under Grant No. 10475107 and 10505007.

\end{document}